\documentclass[letterpaper,english,aps,prl,twocolumn,notitlepage]{revtex4}
\usepackage[T1]{fontenc}
\usepackage{amsmath}
\usepackage{amssymb}
\usepackage{graphicx}
\usepackage{soul}
\usepackage{comment}
\usepackage{cancel}

\makeatletter

\pdfpageheight\paperheight
\pdfpagewidth\paperwidth

\usepackage{color}
\usepackage{dsfont}
\usepackage{dsfont}

\usepackage[usenames,dvipsnames,svgnames,table]{xcolor}

\definecolor{violet}{rgb}{0.58, 0.0, 0.83}

\makeatother

\begin{document}

\title{Counting edge modes via dynamics of boundary spin impurities}

\author{Umar Javed$^{1}$, Jamir Marino$^{2}$, Vadim Oganesyan$^{3,4}$, Michael Kolodrubetz$^1$}
\affiliation{$^1$Department of Physics, The University of Texas at Dallas, Richardson, Texas 75080, USA}
\affiliation{$^2$Institut f\"{u}r Physik, Johannes Gutenberg-Universit\"{a}t Mainz, D-55099 Mainz, Germany}
\affiliation{$^3$Physics program and Initiative for the Theoretical Sciences, The Graduate Center, CUNY, New York, NY 10016, USA}
\affiliation{$^4$Department of Physics and Astronomy, College of Staten Island, CUNY, Staten Island, NY 10314, USA}

\begin{abstract}
We study dynamics of the one-dimensional Ising model in the presence of static symmetry-breaking boundary field via the two-time autocorrelation function of the boundary spin. We find that the correlations decay as a power law. We uncover a dynamical phase diagram where, upon tuning the strength of the boundary field, we observe distinct power laws that directly correspond to changes in the number of edge modes as the boundary and bulk magnetic field are varied. We suggest how the universal physics can be demonstrated in current experimental setups, such as Rydberg chains.
\end{abstract}

\maketitle

The interplay of many-body interactions and correlations~\cite{mahan20089,giamarchi2003quantum} lays at the foundations of emergent phenomena  from condensed matter and atomic and molecular systems to high energy physics. 
The challenge posed by treating inter-particle interactions non-perturbatively in extended systems suggests that one should search for simpler setups to serve as stepping stones towards increasingly complex problems. 
An archetypal class of such systems are quantum impurity models in strongly correlated systems. By embedding one or a few degrees of freedom in a many-body medium, one can often treat strong impurity-environment couplings exactly, with the goal of building understanding and applying it towards even more complex scenarios.
Examples include the Anderson orthogonality catastrophe in local quenches of gapless systems~\cite{nozieres1969singularities,anderson1967infrared, silva2008statistics,cetina2016ultrafast,knap2012time}, interaction-dependent transport in one-dimensional junctions~\cite{kane1992transport,kane1995impurity}, the build-up of entanglement among magnetic impurities and their surrounding fermionic or bosonic environments ~\cite{latta2011quantum, cox1998exotic, andrei1995}, and the formation of  polarons in solid state systems or cold atomic clouds~\cite{devreese2009frohlich,emin2013polarons,schirotzek2009observation,jorgensen2016observation,fukuhara2013quantum,wenz2013few}. 
\begin{figure}[b]
    \centering
    \includegraphics[width=0.85\columnwidth]{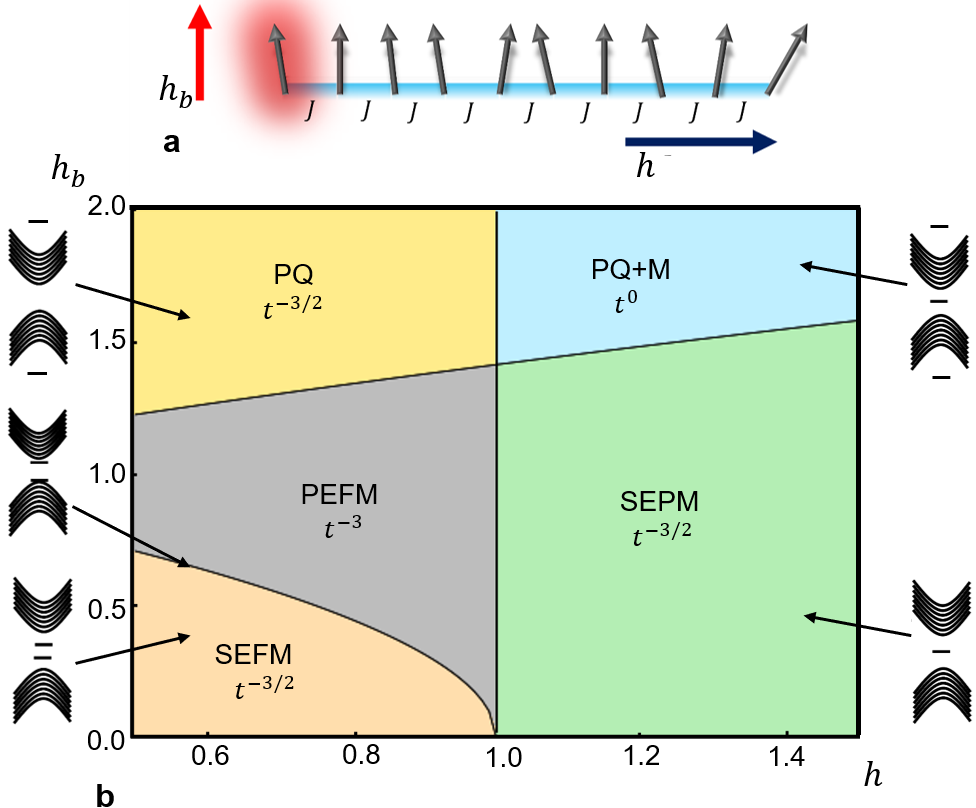}
    \caption{We study magnetization dynamics of the edge spin in the 1D transverse field Ising model in the presence of boundary field $h_b$, resulting in power law relaxation of the boundary autocorrelation function (Eq.~\ref{eq:autocorrelation_function_definition}) as a function of $h_b$ and transverse field ($h$). Phase transitions between different power laws correspond to a change in the number of edge modes. Phases are labeled according to the behavior of their edge mode: soft-edge para/ferromagnetic (SEPM/SEFM) for cases where the edge mode relaxes slowly as $t^{-3/2}$, pinned edge ferromagnetic (PEFM) where the edge is held fixed by the field, and protected qubit (PQ) plus Majorana (+M) where large boundary field creates protected edge modes.}
       \label{fig:edge states}
\end{figure}

This work aims to examine new aspects of a quintessential impurity model hosting edge modes, namely the one-dimensional interacting Ising chain with a strong symmetry-breaking boundary field. 
The impact of boundary fields on the critical point of an extended system is a subject of active interest both in classical and quantum statistical mechanics~\cite{PhysRevB.96.241113,Vasseur,Francica,Prev_2,Kun,Prev_4,Campostrini}. 
Although at leading order impurities appear to be a sub-leading correction in a large system, RG-relevant boundary perturbations can actually induce the formation of new phases and dictate the onset of  critical exponents~\cite{hw1997}.   In addition to its fundamental importance, the response of bulk systems to relevant boundary perturbations can yield novel edge modes, when in turn have the potential to be utilized as a   resource  in quantum computing.  
By uncovering unexpected boundary dynamics in a well-studied model, this work should open the door for further extending our understanding of universal non-equilibrium phenomena in strongly interacting quantum systems.

\emph{Model -- }
We  consider a 1D transverse field Ising model (TFIM) with in the presence of local boundary field with Hamiltonian 
\begin{equation} \label{ising}
    H=-J\sum_{n=1} ^{L-1} \sigma_{n} ^z \sigma_{n+1} ^z - h\sum_{n=1} ^{L} \sigma_n ^x-h_b \sigma_1 ^z
\end{equation}
where $\sigma_j ^{(z,x)}$are Pauli matrices, $J=1$ is the exchange interaction, $h$ is the transverse field, and $h_b$ is a static boundary field along the  $z-$direction. In the absence of boundary field $(h_b=0)$, the TFIM  has $\mathds{Z}_2$ symmetry and undergoes a continuous phase transition at $h=J$, separating the ferromagnetic phase $(h<J)$ and paramagnetic phase $(h>J)$. 

\begin{figure*}
    \centering
    \includegraphics[width=0.9\textwidth]{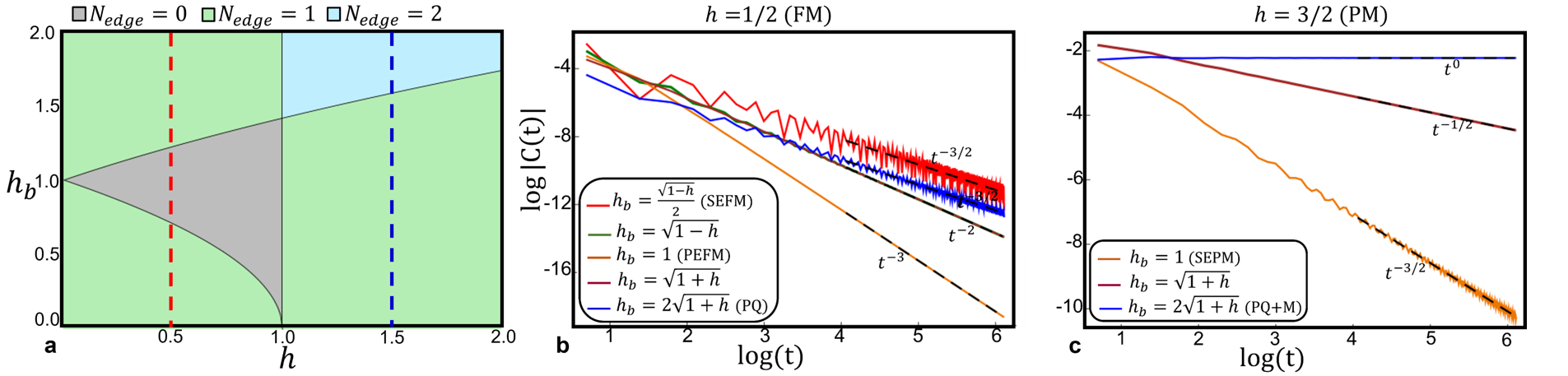}
    \caption{(a)Phase diagram representing the edge states in the presence of boundary field $h_b$.  (b,c)Plots of autocorrelation function for $h=0.5$ and $h=1.5$ with different values of $h_b$ taken across the different phases of edge states represented by red and blue lines in the phase diagram, respectively. Power laws are shown as guides to the eye and match those shown in Fig.~\ref{fig:edge states}.
    }
       \label{fig:correlation function}
\end{figure*}

\emph{Numerical results --}
Motivated by the search for dynamical probes of edge modes~\cite{Vasseur}, we now focus on the dynamics of the boundary spin. 
Specifically, we calculate the connected autocorrelation function of the boundary spin's magnetization,
\begin{equation}
    C(t)=\langle \sigma_1 ^z (0) \sigma_1 ^z (t) \rangle - \langle \sigma_1 ^z (0) \rangle \langle \sigma_1 ^z (t) \rangle
    \label{eq:autocorrelation_function_definition}
\end{equation}
in the many-body ground state.  In the absence of a boundary field, $C(t)$ has been found to decay as a power law, $C(t) \propto t^{-\alpha}$, near criticality ~\cite{ACF}. We find that a power law persists in the presence of $h_b$, but the exponent $\alpha$ is modified. As shown in Fig.~\ref{fig:correlation function}, critical lines emerge in which the power $\alpha$ changes sharply. On the critical lines between these boundary ``phases of matter,'' other power laws emerge.
The goal of the remainder of this paper will be to understand these emergent power laws in the dynamical response.

\emph{Edge states -- }
Previous works on the TFIM with either longitudinal or transverse boundary field have demonstrated the existence of edge states via the Jordan-Wigner mapping to free Majorana fermions ~\cite{Francica,Prev_2,Kun,Prev_4,Campostrini}. To study the connection between these edge states and the boundary spin dynamics, we perform a
Jordan-Wigner transformation, given by ~\cite{William}
\begin{equation} 
    \sigma_n ^x =i \eta_{A,n} \eta_{B,n},\ \sigma_n ^z =i \gamma \left(\prod_{j=1} ^{n-1} i \eta_{A,j} \eta_{B,j}\right) \eta_{A,n}
\end{equation}
where an ancilla Majorana $\gamma=\gamma^\dagger$ is added to the usual Jordan-Wigner string such that the boundary field also maps to a Majorana hopping term: 
\begin{equation} \label{majorana chain}
        H=-iJ\sum_{n=1} ^{L-1} \eta_{B,n} \eta_{A,n+1} -i h\sum_{n=1} ^{L} \eta_{A,n} \eta_{B,n} -i h_b \gamma \eta_{A,1}
\end{equation}

In the absence of $h_b$, Eq.~\ref{majorana chain} represents the Kitaev chain ~\cite{kitaev2001unpaired}, which has a topological Majorana zero mode on the FM side $(h<J)$. Furthermore, the ancilla Majorana gives a separate (artificial) zero mode, which nevertheless couples into the Kitaev chain for $h_b\neq 0$. In the ferromagnetic phase, the ancillary zero mode gaps out the topological zero mode, yielding a gapped fermion. By contrast, there is no topological zero mode on the paramagnetic side, so the ancillary zero mode remains fixed at $E=0$ despite hybridizing with the Kitaev chain. 

At higher $h_b$, a richer edge state structure emerges, as illustrated in Fig.~\ref{fig:edge states}. For instance, at $h_b=\sqrt{1-h}$ and $h < J$, the edge state merges into the bulk. At $h_b=\sqrt{1+h}$, a second (gapped) edge state emerges out of the top of the band. Analytical expressions for the edge mode wave functions and energies can be found in ~\cite{supplementalmaterial}; they can be exactly solved either for the lattice model or within the low energy field theory. The field theory calculation further supports the idea that the phase transitions at small $h_b$ and $|h-J|\ll 1$ are universal. By contrast, the edge mode which emerges at $h_b = \sqrt{1+h}$ does not show up in the Ising field theory, indicating that it is a non-universal lattice effect. We also note that identical edge states have been found in previous studies of transverse boundary fields ($h\sigma^z_1 \to h_{bx} \sigma^x_1$) upon interchanging $h$ and $J$ ~\cite{Francica}. This comes from the mathematical fact that the transverse boundary field maps to an identical Majorana chain but shifted by one site due to lack of the ancilla Majorana. However, the role of transverse boundary field is non-universal in the Ising field theory; $h_b \sigma^z_1$ is a relevant boundary perturbation with scaling dimension $1/2$, while $h_{bx} \sigma^x_1$ is marginal ~\cite{William,cardy2004boundary}. Therefore, we expect our predictions of symmetry-breaking boundary dynamics to be more robust when taken beyond the clean, non-interacting TFIM.

Crucially, we see that the transition lines where edge modes are gained or lost are precisely the lines where the exponent $\alpha$ dictating edge spin decay changes. This connection between emergent fermionic edge modes and boundary spin dynamics has not previously been explored. We now seek to model the behavior of the boundary spin and explain the origin of this connection.

\emph{Boundary spectral function -- }
To understand the connection between edge modes and boundary dynamics, consider the spectral function 
\begin{equation}
    C(\omega) =2\pi \sum_{n\neq 0} \left| \langle \psi_n | \sigma^z_1 | \psi_0 \rangle \right|^2 \delta \left[ \omega - (E_n - E_0) \right] 
    \label{eq:spectral_function}
\end{equation}
where $|\psi_0\rangle$ is the ground state and $|\psi_n\rangle$ are the excited states. Diagonalizing the Hamiltonian in the fermionic basis,
\begin{equation}
H = \sum_{\ell=1}^{N_{edge}}\epsilon_{\ell}\left(2c_{\ell}^{\dagger}c_{\ell}-1\right)+\sum_{k}\epsilon_{k}\left(2c_{k}^{\dagger}c_{k}-1\right)    
\end{equation}
we can separate the edge ($\ell$) and bulk ($k$) modes in the thermodynamic limit. Note that this solution involves combining the original Majoranas into Dirac fermions $c_k$. This form of $H$ allows for both gapped edge modes and Majorana zero modes, for which $c_\ell = \gamma_{edge}$ with $\epsilon_\ell=0$. One can choose $\epsilon \geq 0$ for all modes, such that $|\psi_0\rangle$ is the vacuum state of the $c$-fermions. Then we immediately see that, since $\sigma^z_1=i\gamma \eta_{A1}$ is a 2-fermion operator, $|\psi_n\rangle$ is restricted to states with two fermion excitations above the vacuum, $|\psi_n\rangle = c_\alpha^\dagger c_\beta^\dagger |\psi_0\rangle$ in order for the matrix element not to vanish. Going back to $C(t)$, we have
\begin{equation}
    C(t) = \frac{1}{2} \sum_{\alpha \neq \beta} e^{-2i(\epsilon_\alpha + \epsilon_\beta)t} \underbrace{\left| \langle \psi_0 | c_\alpha c_\beta \sigma^z_1 | \psi_0 \right|^2}_{f_{\alpha \beta}}
    \label{eq:autocorrelation_function_fermions}
\end{equation}
where $\alpha,\beta$ iterate over edge and bulk modes.

At late times, we can solve Eq.~\ref{eq:autocorrelation_function_fermions} via a saddle point approximation. There are three separate situations to consider:
\begin{enumerate}
    \item If $\alpha$ and $\beta$ are both edge states, which is possible for $N_\mathrm{edge}\geq 2$, then one has infinitely long-lived oscillations proportional to $\cos\left[2(\epsilon_\alpha + \epsilon_\beta)t\right]$ as long as the matrix element $f_{\alpha\beta}$ is of order one, as expected for edge states. 
    \item If $\alpha$ is an edge state and $\beta=k$ is a bulk state, then in the thermodynamic limit we can replace $\sum_k \to (L/2\pi) \int dk$. For $t \gg 1/J$, this integral is dominated by the saddle points of the fast-oscillatory term which, for the bulk TFIM, are at $k=0$ and $k=\pi$. As shown ~\cite{supplementalmaterial}, this matches the numerically found exponent $t^{-3/2}$ if the matrix element scales as $f_{\alpha k} \sim k^2$. Such a scaling emerges naturally in the field theory limit from the bulk modes with open boundary conditions, whose (Majorana) wave functions are proportional to $\sin(k) \sim k$ at low momentum. This power law decay is an envelope for $e^{-2i\epsilon_\alpha t}$ oscillations due to the edge mode.
    \item If $\alpha=k$ and $\beta=k^\prime$ are both bulk states, then the sum becomes and integral over $k$ and $k^\prime$. Assuming separability of $f_{k k^\prime} \sim k^2 k^{\prime 2}$, we find $C(t) \sim \left(t^{-3/2}\right)^2 = t^{-3}$, as seen numerically.
\end{enumerate}
For $N_{edge}=0$, only case 3 is possible, while cases 2 and 3 are possible for $N_\mathrm{edge}=1$. However, the late time dynamics will be dominated by the slowest decaying exponent, leading to the prediction $|C(t)| \sim t^{-(3/2) (2 - N_\mathrm{edge})}$ as seen in Fig.~\ref{fig:correlation function}.

\emph{Boundary phases of matter -- }
Having established the existence of edge states and their connection to the edge spin dynamics, $C(t)$, we now discuss the physical meaning of these power law decays and provide labels for the boundary ``phases of matter.'' Let's start with the low-field limit, $h_b \ll 1$, for which the physics near the critical point are universal. In this regime, there are three phases of matter, which we now discuss in detail:
\begin{itemize}
    \item \emph{Soft-edge paramagnet (SEPM)} ($h > J$, $h_b < \sqrt{1+h}$): This is the phase extending from the $h_b=0$ paramagnet in which a Majorana zero mode persists, causing slow $t^{-3/2}$ relaxation of the edge magnetization. Perturbing away from the critical point at $h=J$ and $h_b = 0$, one can think of this phase as where the bulk mass gap $\sim h - J$ is more relevant than the boundary perturbation, which corresponds to an energy scale $E_b \sim h_b^2$ ~\cite{William}. Since symmetry-breaking field is not important in defining the paramagnetic phase, the boundary dynamics of the SEPM is smoothly connected to the conventional paramagnet at $h_b=0$.
    \item \emph{Soft-edge ferromagnet (SEFM)} ($h < J$, $h_b < \sqrt{1-h}$): This is the phase extending up from the $h_b=0$ ferromagnet in which the ancilla Majorana couples to the topological edge Majorana and opens a gap, again causing slow $t^{-3/2}$ relaxation of the edge magnetization. In the spin language, this corresponds to a finite gap between the symmetry-breaking ground states which is proportional to the symmetry-breaking field $h_b$. This destruction of spontaneous symmetry breaking results in an increase in the edge spin relaxation from the ferromagnet, for which it must decay to a constant: $|C(t)| \sim t^0$ as $t\to\infty$ for $h_b=0$. From a field theory perspective, this is the phase as where the symmetry breaking  mass gap $\sim J - h$ is more relevant than the boundary perturbation. However, unlike the SEPM, the soft-edge ferromagnet is not smoothly connected to the $h_b=0$ ferromagnet because the symmetry-breaking boundary field fundamentally changes the symmetry-breaking ferromagnetic phase.
    \item \emph{Pinned-edge ferromagnet (PEFM)} ($h < J$, $\sqrt{1-h} < h_b < \sqrt{1+h}$): This is the phase in which all edge modes have merged into the bulk, resulting in fast $t^{-3}$ relaxation of the edge magnetization. In the spin language, this corresponds to a case where one of the original symmetry breaking ground states, namely $|\Downarrow\rangle$, has merged into the bulk continuum, meaning that single itinerant domain wall excitations become less costly than a global flip of the Ising spins. In this case, the bulk (and edge) are pinned to a single ground state, removing any meaningful notion of symmetry breaking at the boundary \footnote{Note that there is still a meaningful notion of symmetry breaking in the bulk, as the dynamics of a large domain wall in the bulk of the system that is oriented opposite to the boundary field is, nevertheless, infinitely long-lived in the infinite domain limit.}. Field theoretically, this is the phase where the boundary perturbation becomes the dominant scale, being more relevant than the mass gap $\sim J - h$. Morally, this phase of matter bears resemblance to the fixed boundary condition case of boundary CFT ~\cite{William}, but with the important caveat that the bulk is weakly gapped in a symmetry-breaking fashion.
\end{itemize}
A useful analogy for thinking of these low-field phases of matter is that the pinned-edge ferromagnet is the (boundary) quantum critical fan emerging from the bulk critical point, where boundary field plays the role of temperature. The shape of the fan is dictated by boundary exponents, notably $z_b = 2$ from $t_b \sim h_b^{-2}$. Unlike conventional pictures of the quantum critical fan, however, there are phase transitions between the boundary dynamics in the different phases, rather than  crossovers. 

While the high-field phases of matter for $h_b > \sqrt{1+h}$ are not universal, in the sense that they come from high momentum lattice physics that is not present in the Ising field theory, they are nevertheless robust within this lattice model. The key point in both phases is that a fermionic edge state emerges out of the top of the single particle band. For $h_b \to \infty$, this can be thought of as the edge qubit, which is in a large magnetic field, $h_b \sigma^z_1$. The question is then how this edge qubit is dressed by excitations of the bulk continuum. For $h_b > \sqrt{1+h}$, the edge spin hybridizes with the bulk, but remains stable. For $h_b < \sqrt{1+h}$, bulk domain walls hybridize with the edge qubit and destabilize it. Therefore, we refer to these phases of matter as the protected qubit (PQ, $h < J$) and protected qubit + Majorana (PQ+M, $h>J$) to reflect the fact that the edge Majorana remains stable for $h > J$ as well. It is particularly notable that the edge correlation function asyptotes to oscillate with finite amplitude ($|C(t)| \sim t^0$) within the PQ+M phase, reflecting the fact that both a fermion and Majorana edge mode coexist, both of which are excited by the $\sigma^z_1$ operator.

\emph{Experimental realizations -- }
Recent experimental advances have made it possible to simulate spin systems in a well-controlled manner. A particularly well-developed platform to explore the physics studied here is with kinetically constrained spin models as realized in tilted Mott insulators of bosons ~\cite{simon2011quantum, Greiner} or, more recently, one- and two-dimensional arrays of Rydberg atoms ~\cite{Exp_1,Exp_2,Exp_3,Greiner,Exp_5}. In Rydberg atoms, the ground state $| g\rangle$ and Rydberg state $|r \rangle$ of the atom can be mapped to a spin 1/2 by considering $| g\rangle = |\uparrow \rangle$ and $| r\rangle = |\downarrow \rangle$. Adding strong dipole-dipole interactions between the Rydberg atoms gives a Hamiltonian 
\begin{equation}
    H= \hbar \Omega  \sum_i \sigma_i ^x - \sum_i \Delta \sigma_i ^z  + \sum_{i \neq j} V_{ij} \sigma_i ^z \sigma_j ^z
\end{equation}
where $\Omega$ is the Rabi frequency of an external drive and $\Delta$ is its detuning frequency, both of which in principle can be controlled locally. The interactions can be made sufficiently strong that the no nearest neighbors can simultaneously be in the Rydberg state, which enables an antiferromagnetic ground state that breaks $\mathds{Z}_2$ symmetry. This model is in the 1D Ising universality class, can be realized in its ground state, can be locally controlled, and has the nice property that a boundary $\sigma^z$ field acts precisely as the symmetry-breaking field required above.

The universal boundary dynamics at low $h_b$, namely the SEPM, SEFM, and PEFM phases, will be accessible in this Rydberg model. We propose two routes to measure the relevant dynamics. First, $C(t)$ can be measured directly using the Hadamard test by directly coupling the boundary spin to an ancilla qubit such that the boundary autocorrelation function maps to coherence of the ancilla ~\cite{Googleexpt}. Second, experimentalists could instead measure $\sigma^z_1$, time evolve, and then measure again, which is in principle possible in these Rydberg tweezer arrays. Finally, we note that although the large $h_b$ phases are not universal, other similar boundary dynamics may be realized for the Rydberg model in the presence of large, symmetry-breaking boundary field.

\emph{Conclusion -- }
In conclusion, we have uncovered an unexpected dynamical signature of emergent edge states in the transverse field Ising model with symmetry-breaking boundary field. Despite sharing a common origin with well-studied effects such as dynamics in boundary conformal field theories (bCFT) ~\cite{William,berdanier2019universal,calabrese2005evolution,calabrese2006time,calabrese2016quantum,calabrese2007entanglement} or boundary phase transitions (e.g. wetting transitions) ~\cite{Campostrini, Kun}, these edge dynamics have a distinct signature. We show that the dynamics are universal at low boundary field and, while the high field regime is not universal, it is nevertheless likely to emerge in similar lattice models due to its simple physical origin.

These phase transitions in the edge dynamics due to it's transparent physical mechanism open a number of questions, such as how they extend to other phases and phase transitions, for instance those whose low-energy dynamics is well-described by a bCFT.As a universal property of the Ising field theory, these effects will survive RG-irrelevant integrability breaking interactions in the bulk . 
The existing results can certainly be immediately extended to a wider class of observables, such as the fidelity susceptibility with regards to the boundary field, which can be seen to diverge in the $t^{-3/2}$ and $t^0$ based on its expression as an integral over 2-time correlations ~\cite{polkovnikov2011universal,zanardi2006ground}. Knowing to look for such a dynamical signature, we postulate that sharp dynamical phase transitions will be readily found on re-examining a host of other well-studied near-critical systems with boundary perturbations. Similar interesting edge dynamics may appear when the static boundary field is replaced by dephasing, for which it may be thought of as a manifestation of the many-body Zeno effect \cite{berdanier2019universal,froml2019fluctuation,dolgirev2020non}. Therefore, our work opens up a new window into universal non-equilibrium boundary phenomena, which should continue to be explored for both fundamental physics and as a potential platform for protecting quantum information.

\emph{Acknowledgments -- }The authors acknowledge useful discusssions with Ehud Altman, Sarang Gopalakrishnan,Romain Vasseur, Tarun Grover, and Aditi Mitra. J.M. acknowledges support from the Dynamics and Topology Centre funded by the State of Rhineland Palatinate and the Deutsche Forschungsgesellschaft (DFG) through the grant HADEQUAM-MA7003/3-1. Work by M.H.K. and U.J. was performed with support from the National Science Foundation through award number DMR-1945529 and the Welch Foundation through award number AT-2036-20200401. Part of this work was performed at the Aspen Center for Physics, which is supported by National Science Foundation grant PHY-1607611, and at the Kavli Institute for Theoretical Physics, which is supported by the National Science Foundation under Grant No. NSF PHY-1748958. We used the computational resources of the Lonestar 5 cluster operated by the Texas Advanced Computing Center at the University of Texas at Austin and the Ganymede and Topo clusters operated by the University of Texas at Dallas' Cyberinfrastructure and Research Services Department.

\bibliography{biblography.bib}

\onecolumngrid

\section*{Supplementary information}

In this supplement, we provide additional analytical calculations supporting the results in the main text.

\section{Analytical calculations of edge states}
In this section we provide analytical solution of edge states for 1D transverse field Ising model (TFIM) in the presence of boundary field $h_b$, in the lattice and field theory limits. The Hamiltonian for our model has the form
\begin{equation} \label{isingA}
    H=-J\sum_{n=1} ^{L-1} \sigma_{n} ^z \sigma_{n+1} ^z - h\sum_{n=1} ^{L} \sigma_n ^x-h_b \sigma_1 ^z
\end{equation}
\subsection*{Lattice model approach}
In order to study the edge states of our model, we first map our Hamiltonian (Eq.~\ref{isingA}) into Majorana fermions using the Jordan-Wigner transformation.
The final form of our Hamiltonian after the Jordan-Wigner transformation is
\begin{equation} \label{majorana chainA}
        H=-iJ\sum_{n=0} ^{L-1} \eta_{B,j} \eta_{A,n+1} -i h\sum_{n=0} ^{L} \eta_{A,n} \eta_{B,n} -i h_b \gamma \eta_{A,1}
\end{equation}
where we consider $J=1$, each site $j$ is split up into Majoranas $\eta_{Aj}$ and $\eta_{Bj}$, and $\gamma$ is the ancilla Majorana  used to write our boundary term as hopping term in Majorana fermions.
Assuming $h_{b}\neq0$ so that any edge states must include hybridization with the ancilla Majorana $\gamma$. We postulate unnomralized edge state of the form
\begin{equation}
    \tilde{c}=\gamma+\sum_{j}\left(\alpha_{j}\eta_{Aj}+\beta_{j}\eta_{Bj}\right)
\end{equation}
where $\alpha_{j}=iAr^{j-1}$, $\beta_{j}=Br^{j-1}$ and $|r|<1$ is required to get well defined edge state.

Edge states satisfy the commutation relation of the form
\begin{equation}\label{commutatorA}
    [H,\tilde{c}]=E\tilde{c}.
\end{equation}
After calculating the commutator and equating the coefficients to solve for $E$ and $\tilde{c}$ we get,
\begin{align}
E & =\pm\frac{2\sqrt{h_{b}^{2}\left(1-h^{2}\right)-h_{b}^{4}}}{\sqrt{1-h_{b}^{2}}}\\
r & =\frac{h}{1-h_{b}^{2}}\\
A & =\frac{E}{2h_{b}}\\
B & =\frac{hh_{b}}{1-h_{b}^{2}}
\end{align}
For small $h_{b}\ll1$, we can immediately see why $h<1$ is required
to give a real energy. In this regime, notice that $r>0$, which implies
that the modes on each sublattice decay without oscillatory sign,
i.e., they happen near zero momentum. This implies that the results
in this regime will be captured by field theory. By contrasts, for
$h_{b}>1$, $r<0$, in which case the modes are near $k=\pi$ and
thus should not be expected to be universal.

Similarly solving  Eq.~(\ref{commutatorA}) for zero mode $E=0$ we get $A=0$ and 
\begin{align}
B & =\frac{h_{b}}{h}\\
r & =\frac{1}{h}
\end{align}
Thus we see why the extra zero mode only exists for $h>1$; otherwise,
it doesn't satisfy the crucial condition $|r|<1$.

\subsection*{Field theory approach}

Having solved the lattice model exactly, we now solve for the edge modes in the field theory formulation to more precisely demonstrate universality. A straightforward derivation of the field theory is obtained by a simple Taylor expansion.
Starting from the lattice model,
\begin{equation}
H=-ih_{b}\gamma\eta_{A0}-ih\sum_{j=0}^{\infty}\eta_{Aj}\eta_{Bj}-iJ\sum_{j=0}^{\infty}\eta_{Bj}\eta_{A,j+1}
\end{equation}
we can replace on-site Majoranas by fields: $\eta_{\sigma,j}\to\eta_{\sigma}(x=ja)$
where $a$ is the lattice spacing. Then 
\begin{equation}
\eta_{A,j+1}\approx\eta_{A}(x)+a\partial_{x}\eta_{A}(x)
\end{equation}
where $x=ja$. Replacing the sum by an integral, the Hamiltonian becomes
\begin{align}
H & \approx-ih_{b}\gamma\eta_{A}\left(0\right)-i\int_{0}^{\infty}dx\left[h\eta_{A}(x)\eta_{B}(x)+J\eta_{B}(x)\left(\eta_{A}(x)+a\partial_{x}\eta_{A}(x)\right)\right]\\
 & =-i\lambda\gamma\eta_{A}(0)-i\int_{0}^{\infty}dx\left[\Delta\eta_{B}(x)\eta_{A}(x)+c\eta_{B}(x)\partial_{x}\eta_{A}(x)\right]
\end{align}
where $\Delta\propto J-h$, $\lambda\propto h_{b}$, and we have used
that the $A$ and $B$ Majorana fields anticommute. 
Based on the lattice solution, we postulate the edge state solution
\begin{equation}
\tilde{c}=\gamma+\int_{0}^{\infty}dx\left(A\eta_{A}(x)+B\eta_{B}(x)\right)e^{-\kappa x}
\end{equation}
Again we calculate commutators and equate coefficients for $E\neq 0$, assuming the normalization $\left\{ \eta_{\sigma}(x),\eta_{\sigma^{\prime}}(x^{\prime})\right\} =\delta_{\sigma\sigma^{\prime}}\delta(x-x^{\prime})$. We get
\begin{align}
 A & =\frac{iE}{\lambda} \\
 B & =-\frac{2\lambda}{c} \\
-iB\left(c\kappa-\Delta\right) & =EA \label{EA_field_theory}\\
-iA\left(c\kappa-\Delta\right) & =EB \\
\kappa & =\frac{\Delta}{c}-\frac{2\lambda^{2}}{c^{2}} \\
E^{2} & =\frac{4\lambda^{2}}{c^{2}}\left(c\Delta-\lambda^{2}\right)
\end{align}
Note that $E^{2}$ can be positive at small enough $\lambda^{2}<c\Delta$ and we see emergent edge states on the FM side, this shows
that the low $\lambda$ edge states are consistently determined within the field theory. Note that there is no re-emergent edge mode at large $\lambda$ within the field theory, as expected from our previous calculation that these correspond to $k\sim\pi$.

Finally, let's consider the zero mode $E=0$. In that case, as before, $A=0$
and from Eq.~\ref{EA_field_theory}, we get
\begin{equation}
\kappa=\frac{\Delta}{c}
\end{equation}
\section{Spectral function}
Consider the connected boundary spin autocorrelation function for
the Ising model with boundary field:
\begin{equation}
C\left(t\right)=\langle\psi_{0}|\sigma_{1}^{z}(0)\sigma_{1}^{z}(t)|\psi_{0}\rangle-\langle\psi_{0}|\sigma_{1}^{z}|\psi_{0}\rangle^{2}
\end{equation}

where $|\psi_{n}\rangle$ denotes the $n$th many body energy eigenstate
with energy $E_{n}$ (ground state is $E_{0}$). Inserting an identity
and Fourier transforming, we have
\begin{align}
C\left(t\right) & =\sum_{n}\langle\psi_{0}|\sigma_{1}^{z}(0)|\psi_{n}\rangle\langle\psi_{n}|\sigma_{1}^{z}(t)|\psi_{0}\rangle-\langle\psi_{0}|\sigma_{1}^{z}|\psi_{0}\rangle^{2} \label{corrA}\\
C(\omega) & =2\pi\sum_{n>0}\delta\left(E_{n}-E_{0}-\omega\right)\left|\langle\psi_{0}|\sigma_{1}^{z}|\psi_{n}\rangle\right|^{2} 
\end{align}
Notice a few things about this expression:
\begin{enumerate}
\item $C(\omega)$ is positive and real.
\item $C(\omega)$ is only nonzero for $\omega>0$ in the case of a gap.
If gapless, then one instead has $\omega\geq0$.
\end{enumerate}

Noting that $\sigma_{1}^{z}=i\gamma \eta_{A1}$
is a fermion bilinear, we have to consider the excited states  of the form 
\[
|\psi_{n}\rangle=c_{\alpha}^{\dagger}c_{\beta}^{\dagger}|\psi_{0}\rangle.
\]
Each such pair will correspond to a peak at 
\begin{equation}
\omega=E_{n}-E_{0}=2\left(\epsilon_{\alpha}+\epsilon_{\beta}\right).
\end{equation}
We can write Eq.~\ref{corrA} as,
\begin{equation}\label{corrA1}
    C(t) =\frac{1}{2}\sum_{\alpha \neq \beta }e^{2i\left(\epsilon_{\alpha}+\epsilon_{\beta}\right)t}\underbrace{\left|\langle\psi_{0}|c_{\alpha}c_{\beta} \sigma_1 ^z|\psi_{0}\rangle\right|^{2}}_{f(\alpha,\beta)}
\end{equation}
\subsection*{Emergence of power laws}

If we approximate the Hamiltonian by a countable set of edge modes
and a bulk continuum whose properties are essentially similar to the
case of periodic boundary conditions, then we get
\begin{equation}
H\approx\sum_{\ell=1}^{N_{edge}}\epsilon_{\ell}\left(2c_{\ell}^{\dagger}c_{\ell}-1\right)+\sum_{k}\epsilon_{k}\left(2c_{k}^{\dagger}c_{k}-1\right)
\end{equation}
where $k=2\pi n/L$ are the momentum modes. Notice that zero modes
are accomodated here by simply having $\epsilon_{\ell}=0$ and a single
zero mode is counted the same as a pair of gapped edge modes (i.e.,
the case $h\gg1$ and $h_{B}\gg1$ would be counted as $N_{edge}=2$
while $h\ll1$ and $h_{B}\ll1$ would be $N_{edge}=1$). In the limit
$L\to\infty$, the spectral function can be written as a sum over
three types of term:
\begin{enumerate}
\item If $N_{edge}\geq2$, then there would be terms where we sum over all pairs of edge modes $\alpha(\ell)\neq\beta(\ell^{\prime})$, with peaks at $\omega=2\left(\epsilon_{\ell}+\epsilon_{\ell^{\prime}}\right)$. Call this contribution $C_{ee}$
\item If $N_{edge}\geq1$, then there would be terms where we sum over all pairs of edge modes $\alpha(\ell)$ and integrate over momenta $\beta(k)$ with peaks at $\omega=2\left(\epsilon_{\ell}+\epsilon_{k}\right)$. Call
this contribution $C_{be}$
\item For all cases, there would be terms
where we integrate over both momenta $\alpha(k)$, $\beta(k')$ with peak at $\omega=2\left(\epsilon_{k}+\epsilon_{k^{\prime}}\right)$. Call this contribution $C_{bb}$
\end{enumerate}
The full $C(t)$ can be obtained by adding these terms.

To get a sense of what time dependence this theory predicts, consider
case 2 of a single edge mode (the responses for multiple edge modes
could simply be summed over). Then using Eq.~\ref{corrA1} $C_{be}(t)$ can be written as
\begin{align}
C_{be}(t) & =\frac{L}{2\pi}\int_{-\pi}^{\pi}dke^{2i\left(\epsilon_{\alpha}+\epsilon_{k}\right)t}\underbrace{\left|\langle\psi_{0}|c_{\alpha}c_{k}\gamma \eta_{A1}|\psi_{0}\rangle\right|^{2}}_{f(\alpha,k)}\\
 & =\frac{L}{2\pi}e^{2i\epsilon_{\ell}t}\int_{-\pi}^{\pi}dke^{2i\epsilon_{k}t}f(\alpha,k) \label{saddleA}
\end{align}
One can calculate the matrix element $f(\alpha,k)$ in the field theory limit,
\begin{equation}
    f(\alpha,k)=\left|\langle\psi_{0}|\gamma \eta_{A1}c_{\alpha}c_{k}|\psi_{0}\rangle\right|^{2}
\end{equation}
using Wicks, theorem we can write:
\begin{equation}
\langle \psi_0| \gamma \eta_{A,1} c_{\alpha} c_k  |\psi_0 \rangle=\langle \gamma \eta_{A,1} \rangle \cancelto{0}{\langle c_{\alpha} ^\dagger c_k ^\dagger \rangle} -\langle \gamma c_{\alpha} ^\dagger \rangle \langle \eta_{A,1} c_k ^\dagger \rangle + \langle \gamma c_k ^\dagger \rangle \langle \eta_{A,1} c_{\alpha} ^\dagger \rangle
\end{equation}
The terms $\langle \gamma c_l ^\dagger \rangle \langle$ and $\langle \eta_{A,1} c_{\alpha} ^\dagger \rangle$ are equal to complex numbers and in the field theory limit, one can easily show that $\langle \eta c_k ^\dagger \rangle \propto \sin(k)$ due to open boundary conditions.

In the late time limit, one can solve Eq.~\ref{saddleA} by the saddle point method.
For the bulk problem, there are two stationary points, $k=0,\pi$,
around which we can expand the energy and $f(\alpha,k)$ and from Eq.~\ref{saddleA} we get $C_{be}(t) \propto t^{-3/2}$. 
Similarly, since for case 3 we don't have any edge contribution, we will integrate over the both momenta $k$ and $k'$ in Eq.~\ref{corrA1} and $f(\alpha,\beta) \propto k^2 k^{\prime 2}$ due to no edge contribution
\begin{align}
C_{be}(t) & =\frac{L^2}{4\pi^2}\int_{-\pi}^{\pi}\int_{-\pi}^{\pi}dk^{\prime} dke^{2i\left(\epsilon_{k}+\epsilon_{k^{\prime}}\right)t}\underbrace{\left|\langle\psi_{k}|c_{\alpha}c_{k^{\prime}}\gamma \eta_{A1}|\psi_{0}\rangle\right|^{2}}_{f(k,k^{\prime})}\\
 & =\frac{L^2}{4\pi^2}\int_{-\pi}^{\pi}\int_{-\pi}^{\pi}dk^{\prime} dke^{2i\left(\epsilon_{k}+\epsilon_{k^{\prime}}\right)t}f(k,k^{\prime}) 
\end{align}

Using the above saddle point approximation, one can easily solve above integral and show that $C_{bb}(t) \propto t^{-3}$. Finally, for case 1 $f(\alpha,\beta)$ will be of order 1 and from Eq.~\ref{corrA1} one can easily see that $C_{ee}(t)$ will not decay and will oscillate forever.


%
\end{document}